\UseRawInputEncoding
\documentclass[preprint,12pt]{elsarticle}
\usepackage{amsmath}
\usepackage{amssymb}
\usepackage{color}
\usepackage{ifsym}
\usepackage{lineno}
\usepackage{siunitx}
\usepackage{wasysym}
\usepackage{pifont}

\journal{Nuclear Instruments and Methods in Physics Research A}

\newcommand{\gbar}{\mbox{$\mathrm{\bar{g}}$}}

\newcommand{\Pep}{\mbox{$\mathrm{e}^+$}}
\newcommand{\Hbar}{\mbox{$\mathrm{\overline{H}}$}}
\newcommand{\Hbarp}{\mbox{$\mathrm{\overline{H}^+}$}}
\newcommand{\pbar} {\mbox{$\mathrm{\overline{p}}$}}
\newcommand{\Ps} {\mbox{$\mathrm{Ps}$}}
\definecolor{Cerulean}{rgb}{0.0, 0.48, 0.65}
\definecolor{forestgreen}{rgb}{0.13, 0.55, 0.13}
\definecolor{limegreen}{rgb}{0.2, 0.8, 0.2}
\definecolor{tangerineyellow}{rgb}{1.0, 0.8, 0.0}
\definecolor{saffron}{rgb}{0.96, 0.77, 0.19}
\DeclareSIUnit\pos{e^+}

\begin{document}

\begin{frontmatter}
%\preprint{}

\title{Positron accumulation in the GBAR experiment}

\author[add6]{P. Blumer}
\author[add1]{M. Charlton}
\author[add3]{M. Chung}
\author[add4]{P. Clad\'{e}}
\author[add5]{P. Comini}
\author[add6]{P. Crivelli}
\author[add7]{O. Dalkarov}
\author[add5]{P. Debu}
\author[add1]{L. Dodd}
\author[add4,add4_1]{A. Douillet}
\author[add4]{S. Guellati}
\author[add8]{P.-A Hervieux}
\author[add4,add4_1]{L. Hilico}
\author[add4]{P. Indelicato}
\author[add6]{G. Janka}
\author[add10]{S. Jonsell}
\author[add4,add4_1]{J.-P. Karr}
\author[add12,add2]{B. H. Kim}
\author[add11]{E. S. Kim}
\author[add2]{S. K. Kim}
\author[add12]{Y. Ko}
\author[add13]{T. Kosinski}
\author[add14]{N. Kuroda}
\author[add5]{B. M. Latacz\fnref{BL}}
\author[add2]{B. Lee}
\author[add2]{H. Lee}
\author[add12]{J. Lee}
\author[add5]{A. M. M. Leite\fnref{AL}}
\author[add8]{K. L\'{e}v\^{e}que}
\author[add11]{E. Lim}
\author[add5]{L. Liszkay}
\author[add5]{P. Lotrus}
\author[add9]{D. Lunney}
\author[add8]{G. Manfredi}
\author[add5]{B. Mansouli\'{e}}
\author[add13]{M. Matusiak}
\author[add15]{G. Mornacchi}
\author[add16]{V. Nesvizhevsky}
\author[add4]{F. Nez}
\author[add9]{S. Niang\corref{cor1}}
    \ead{samuel.niang@cern.ch}
\author[add14]{R. Nishi}
\author[add6]{B. Ohayon}
\author[add2]{K. Park}
\author[add4]{N. Paul}
\author[add5]{P. P\'{e}rez}
\author[add5]{S. Procureur}
\author[add6]{B. Radics}
\author[add6]{C. Regenfus}
\author[add5]{J.-M. Reymond}
\author[add4]{S. Reynaud}
\author[add5]{J.-Y. Rouss\'{e}}
\author[add4]{O. Rousselle}
\author[add6]{A. Rubbia}
\author[add13]{J. Rzadkiewicz}
\author[add5]{Y. Sacquin}
\author[add17]{F. Schmidt-Kaler}
\author[add13]{M. Staszczak}
\author[add13]{K. Szymczyk}
\author[add14]{T. Tanaka}
\author[add5]{B. Tuchming}
\author[add5]{B. Vallage}
\author[add7]{A. Voronin}
\author[add1]{D. P. van der Werf\corref{cor2}}
    \ead{d.p.van.der.werf@swansea.ac.uk}
\author[add17]{S. Wolf}
\author[add2]{D. Won}
\author[add13]{S. Wronka}
\author[add18]{Y. Yamazaki}
\author[add3]{K. H. Yoo}
\author[add4]{P. Yzombard}
\author{(GBAR collaboration)} 
\author[add1]{C. J. Baker}

\address[add6]{Institute for Particle Physics, ETH Z\"{u}rich, Otto Stern Weg 5, CH-8093 Z\"{u}rich, Switzerland}
\address[add1]{Department of Physics, Faculty of Science and Engineering, Swansea University, Swansea SA2 8PP, UK}
\address[add3]{Ulsan National Institute of Science and Technology, Ulsan 44919, Republic of Korea}
\address[add4]{Laboratoire Kastler Brossel, Sorbonne Universit\'e, CNRS, ENS-Universit\'e PSL, Coll\`{e}ge de France, 4 place Jussieu, 75005 Paris, France}
\address[add5]{IRFU, CEA, Universit\'e Paris-Saclay, F-91191 Gif-sur-Yvette, France}
\address[add7]{Moscow, Russia}
\address[add4_1]{Universit\'e d`{\'E}vry-Val dÕEssonne, Universit\'e Paris-Saclay, Boulevard Fran\c cois Mitterrand, F-91000 \'Evry, France}
\address[add8]{Institut de Physique et Chimie des Mat\'{e}riaux de Strasbourg, 23 rue du Loess, F-67037 Strasbourg, France}
\address[add10]{Department of Physics, Stockholm University, SE-10691 Stockholm, Sweden}
\address[add11]{Department of Accelerator Science, Graduate School, Korea University Sejong Campus, Sejong 30019, Korea}
\address[add12]{Center for Underground Physics, Institute for Basic Science (IBS), Daejeon 34047, Republic of Korea}
\address[add13]{Narodowe Centrum Bada\`n J\c{a}drowych ul. Andrzeja So\l tana 7, 05-400 Otwock, \'Swierk, Poland}
\address[add2]{Department of Physics and Astronomy, Seoul National University, 599 Gwanak-ro, Gwanak-ku, Seoul, 151-742, Korea}
\address[add14]{Institute of Physics, University of Tokyo, 3-8-1 Komaba, Meguro, Tokyo, 153-8902, Japan}
\address[add9]{Universit\'e Paris-Saclay, CNRS-IN2P3-IJCLab, Orsay 91405, France}
\address[add15]{Physics Department, CERN, CH-1211 Geneva 23, Switzerland}
\address[add16]{Institut Laue-Langevin (ILL), 6 rue Jules Horowitz, F-38042 Grenoble, France}
\address[add17]{Johannes Gutenberg Universit\"at, D-55128 Mainz, Germany}

\fntext[BL]{present address: RIKEN, Ulmer Fundamental Symmetries Laboratory, Wako, Saitama 351-0198, Japan.}
\fntext[AL]{present address: Institut Curie, PSL Research University, Radiation Oncology Department, Proton Therapy Centre, Centre Universitaire, 91898, Orsay and  Institut Curie, PSL Research University, Universit\'e Paris-Saclay, CNRS UMR 3347, INSERM U1021, 91898 both in France.}

\cortext[cor1,cor2]{Corresponding authors}
\date{\today}

\begin{abstract}
We present a description of the GBAR positron ($\Pep$) trapping apparatus, which consists of a three stage Buffer Gas Trap (BGT) followed by a High Field Penning Trap (HFT), and discuss its performance. The overall goal of the GBAR experiment is to measure the acceleration of the neutral antihydrogen (\Hbar) atom in the terrestrial gravitational field by neutralising a positive antihydrogen ion (\Hbarp), which has been cooled to a low temperature, and observing the subsequent \Hbar\ annihilation following free fall. To produce one \Hbarp ion, about $10^{10}$ positrons, efficiently converted into positronium (\Ps), together with about $10^7$ antiprotons (\pbar), are required. The positrons, produced from an electron linac-based system, are accumulated first in the BGT whereafter they are stacked in the ultra-high vacuum HFT, where we have been able to trap \num{1.4+-0.2e9}  positrons in 1100 seconds.
\end{abstract}

\end{frontmatter}

%\maketitle %\maketitle must follow title, authors, abstract and \pacs

% References should be done using the \cite and \label commands
\section{Introduction}
\label{Intro}
The aim of the GBAR collaboration \cite{GBAR_Proposal} is to measure the gravitational acceleration (\gbar) of an antihydrogen atom in Earth's gravitational field, following a proposal by Walz and H\"ansch \cite{Walz2004}. The intention, as outlined below, is to form ultra-cold \Hbar\ atoms by employing positive antihydrogen ions, with the anti-ions produced via the two charge exchange reactions:
\begin{align}
\label{eq:Hbar}
\pbar + {\rm Ps} &\rightarrow \Hbar + {\rm e}^- ,\\
\label{eq:Hbarp}
\Hbar + {\rm Ps} &\rightarrow \Hbarp + {\rm e}^-.
\end{align}
\noindent Here the \pbar-\Ps\ collisions produce \Hbar\ atoms, which in their turn can have a further interaction to form the \Hbarp. Subsequently, the anti-ions will be captured and sympathetically cooled in a Paul trap, whereafter they will be neutralised using a pulsed laser and the neutral \Hbar\ will undergo a free fall in the earth's gravitational field. After some time the \Hbar\ atoms strike a surface, resulting in \pbar\ annihilation. The concomitant shower of charged pions will be monitored using position sensitive (micromegas) and time sensitive (fast plastic) detectors, thereby allowing the time between positron removal and pion detection to be derived, and so measure a value for \gbar.

 The~ \Hbarp\ production proceeds via the reactions above by directing a pulse of antiprotons, which originate from the Extra Low Energy Antiproton (ELENA) ring in the Antiproton Decelerator (AD) complex at CERN \cite{Maury_2014}, through a \Ps\ cloud. The \Ps\ atoms are produced by implanting $\Pep$ into a target, manufactured from porous silica, which has a conversion efficiency of about 30\%\ \cite{PhysRevA.81.052703}.  

It is evident from equations (\ref{eq:Hbar}) and (\ref{eq:Hbarp}) that the probability of $\Hbarp$ formation is proportional to the square of the Ps density, which is in turn dependent (in part) upon the positron flux reaching the convertor. It is envisaged \cite{GBAR_Proposal,Perez:1072188} that a 30 ns-wide burst of $\approx 10^{10}$ positrons is needed to produce around one \Hbarp per ELENA \pbar\ bunch. The latter arrives about every 120 seconds, during which time this number of positrons needs to be accumulated, such that the  flux emanating from a positron source needs to be $10^8$-$10^9$ $\Pep$s$^{-1}$, dependent on the efficiency of their accumulation into the system of traps. To reach this goal the GBAR collaboration has chosen a low energy electron linac (9 MeV) for production of the required $\Pep$ beam, as will be explained further in section \ref{sec:PosProd}.

The positrons are to be stored in the HFT which employs a 5 T magnetic field, following initial trapping in the three stage BGT. The first two stages of the BGT accumulate positrons for about 100 ms, whereafter they are stacked in the third stage for around one second (10 stacks), before being moved into the HFT. Once enough positrons have been accumulated, they are axially compressed, expelled and accelerated to an energy of 4 keV. The resulting pulse has a timewidth of 30 ns when reaching the $\Pep$-$\Ps$ convertor target. 

In the following we give a description of the traps and the bunching system, together with experimental data, including the trapping efficiencies at each stage. In section \ref{sec:PosProd} we will briefly describe the linac and the source/moderator setup: more information can be found elsewhere \cite{GBAR_linac}. Section \ref{BGT:Intro} contains an overview of the BGT, while the HFT is described in section \ref{sec:HFT}. Section \ref{sec:conclusions} contains our conclusions and an outlook.

\section{Positron production}
\label{sec:PosProd}

Laboratories using BGTs typically employ a $^{22}$Na positron source followed by a moderator to produce a quasi-monoenergetic beam at low energies. Presently, the strongest commercially available sources have an activity of around 1.85 GBq, where the company iThemba Labs is currently the only supplier. Using this source, a low energy beam with a maximum strength of about $10^7 e^+$s$^{-1}$ \cite{RevSciInst.77.063302} can be produced when used together with a solid neon moderator \cite{doi:10.1063/1.97441}. In conjunction with a BGT, which typically has a capture efficiency of about 20\%, followed by an HFT i n which the positrons have a lifetime of hours, a minimum of 80 minutes would be needed to accumulate the $10^{10}$ positrons required for the GBAR experiment, if there are no further losses in transferring positrons (see sections \ref{subsec:Third_BGT} and \ref{sec:HFT}). Moreover, the half-life of the isotope is 2.6 years, necessitating source replacement every couple of years. The use of nuclear reactors and high energy (18 MeV-180 MeV) electron linacs is not a viable option in the AD hall, due to space, safety and funding constraints, so the GBAR collaboration has chosen a relatively low energy electron linear accelerator to obtain the necessary $\Pep$ production rate.

The linac, manufactured by our collaborators of NCBJ in Swierk (Poland), accelerates electrons to a kinetic energy of 9 MeV in a pulse width of 2.85 $\mu$s with a peak current of 300 mA and a maximum repetition rate of 300 Hz.  Although the linac is able to produce higher energy electrons, which would increase the number of fast positrons, the energy is limited to 9 MeV to prevent activation of the materials in the target and its surroundings. 
The accelerated electrons are implanted into a water-cooled tungsten target, thereby producing high energy gamma rays which in turn lead to positron formation via pair production, and finally a mono-energetic $\Pep$ beam is emitted from a tungsten mesh moderator elevated to a potential of +50 V. The low energy antiparticles are then magnetically guided towards the traps using a system of coils and solenoids. 

The $\Pep$ flux emanating from the transport system is measured using a calibrated NaI(Tl) scintillator-photomultiplier tube arrangement used to record the annihilation gamma rays produced when the beam strikes a target inserted just before the accumulator entrance (for further details see \cite{Basia_Thesis}). 

\begin{figure}[ht]
\centering
\includegraphics[width=\linewidth]{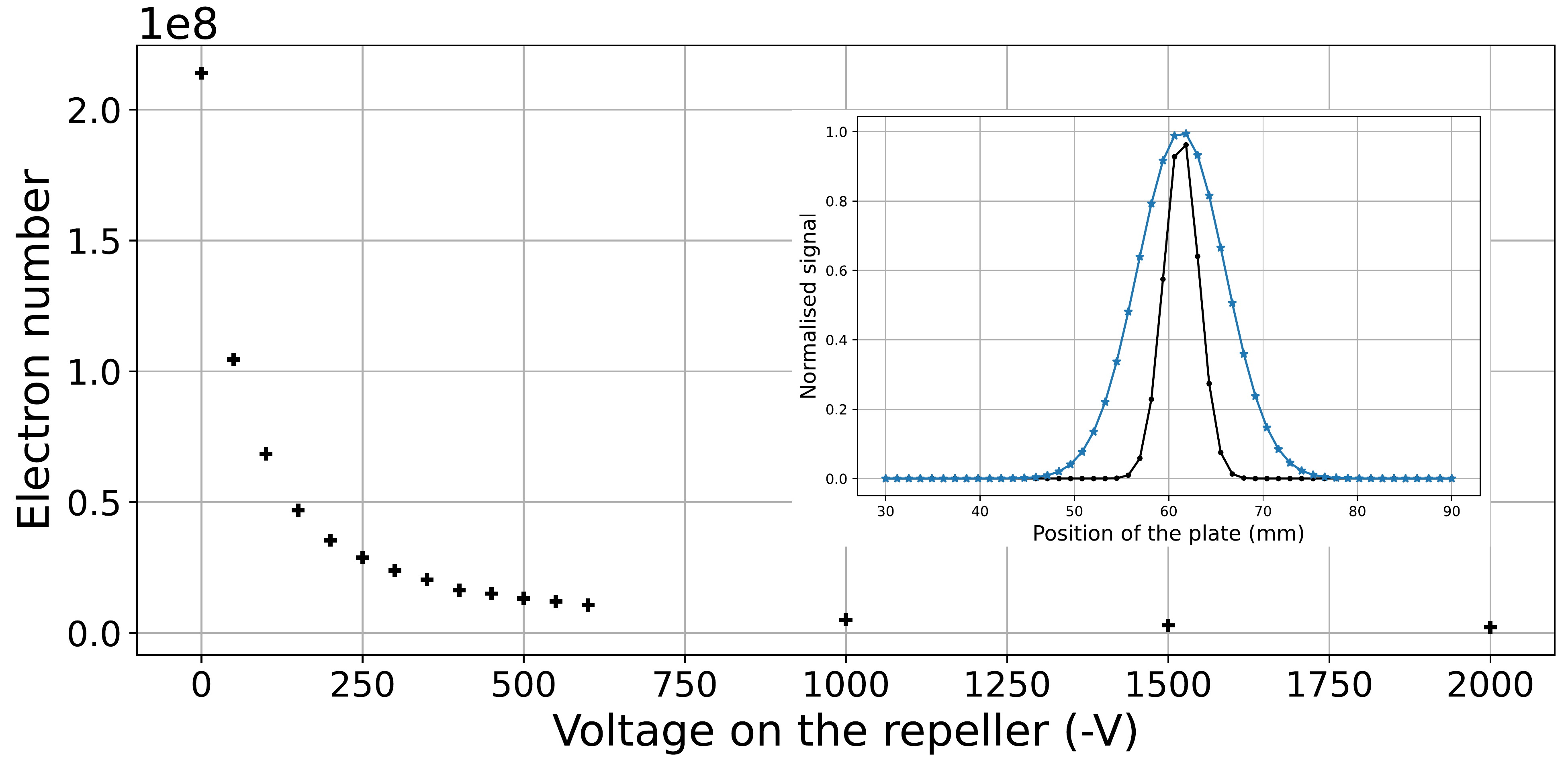}
\caption{The number of electrons per pulse as a function of the retarding grid potential, with the number at 0 V corresponding to $2.14 \times 10^8$. The linac frequency was 2.4 Hz, which was chosen as this is the only frequency which allows staff to be in the GBAR experimental area with the linac in operation. Inset: the vertical widths of the positron ($\bullet$) and electron ({\color{Cerulean}$\bigstar$}) pulses.}
\label{fig:Repel1}
\end{figure}

In addition to the positrons, electrons are also emitted by the target/moderator setup which at low enough kinetic energies will reach the BGT. Their origin is via pair production and as secondaries emitted from the surfaces of the various materials used inside the vacuum chamber. By measuring their current on the diagnostic target as a function of the potential of a retarding grid placed before that target, it was found that the electron kinetic energy ranges from 0 to at least 5 keV (see figure \ref{fig:Repel1}). The number of electrons per linac pulse with the retarding grid on 0 V is 500-1000 times the positron number, while the vertical width of the electron beam (measured by moving the target through the beams) is about twice that of the $\Pep$s (see inset figure \ref{fig:Repel1}). It turns out that the electrons have an effect on the trapping rate, as will be shown in section \ref{sec:First_and_Second}. More information concerning low energy positron beam production using the 9 MeV linac can be found in \cite{GBAR_linac}.

\section{Positron accumulation}
\label{BGT:Intro}
Capturing positrons in Penning-type traps utilising collisions with a buffer gas is a well established method. Pioneered by Surko and co-workers  \cite{PhysRevLett.62.901,PhysRevA.46.5696}, BGTs are nowadays used in many experimental groups around the world (see for instance references \cite{RevSciInst.77.063302,doi:10.1063/1.2221509,doi:10.1063/1.3030774}) and especially in the AD at CERN where the ASACUSA \cite{Kuroda2014}, AEgIS \cite{AGHION201586}, ATHENA \cite{AMORETTI2004679}, ALPHA \cite{AMOLE2014319}, ATRAP \cite{Comeau_2012} and GBAR collaborations have all employed such devices for the initial trapping of the positrons required for their \Hbar\ experiments. Because in all cases the \Hbar\ formation area and experimentation regions need to have ultra-high vacuum (or better) conditions, the BGT is always followed by a HFT.

\subsection{Principles of positron capture in Penning traps with a buffer gas}
\label{BGT:Principle}
The Penning (or Penning-Malmberg) traps used in positron experiments typically consist of a series of cylindrical electrodes immersed in an axial magnetic field. The $\Pep$s are confined in the axial direction by a potential well formed by applying appropriate voltages to the aforementioned electrodes, and in the radial direction by the magnetic field. Depending on the particular experiment, the field can be as low as 0.05 T (as in this work) or up to 2.5 T \cite{Imao_2010}. Gas is admitted into the system and the positrons lose enough energy via collisions such that they are unable to escape from the well.  Experiments have shown that nitrogen gas gives the highest trapping efficiency, typically around 20-30\% \cite{Greaves_1996,RevSciInst.77.063302,10.1088/1361-6455/aba10c}. In the first stage (see section \ref{BGT:Description}) the incoming positrons lose kinetic energy due to electronic excitation of the nitrogen molecules as
\begin{equation}
{\rm e^+ + N_2 \rightarrow e^+ + N^*_2,}
\label{eq:ElecExcite}
\end{equation}
which has a threshold of 8.6 eV. There are, however, competing processes which limit the trapping efficiency, with the main one being positronium formation,
 \begin{equation}
{\rm e^+ + N_2 \rightarrow Ps + N^+_2,}
\label{eq:PsForm}
\end{equation}
which opens at 8.8 eV and results in positron loss.  The relative behaviour of the cross sections for the two processes and the energy width of the incoming $\Pep$ beam are the main physical factors which govern the efficiency of capture into the trap at a given nitrogen gas density. Another process, ionisation of N$_2$, starts at an energy of 15.6 eV but will typically have little effect on the trapping. 

Once the $\Pep$ are trapped, their lifetime is predominantly limited by radial transport, induced by elastic scattering due to the relatively high nitrogen density needed to ensure efficient capture, which leads to their eventual loss on the trap wall. This effect can be counteracted by using a rotating wall electric field (RW) \cite{PhysRevLett.85.1883,ApplSurfSci_194_312,Isaac11}, but the concomitant heating necessitates that an extra gas be used to cool the $\Pep$ at low energies, because N$_2$ has a low cooling rate \cite{GREAVES200290}. Thus, we add CO$_2$ to the second and third stages. When the RW drive has the optimum frequency and amplitude, the radial expansion is halted, or even reversed, such that the lifetime of the cloud is then only limited by annihilation on the gas molecules \cite{NJP_14_075022}. 

It should be noted that at the densities used in the first two stages of the BGT, the $\Pep$ clouds are in the so-called single particle regime. That is, the Debye screening length, $\lambda_D$, is much larger than the particle cloud dimension, where $\lambda_D$ is defined by
\begin{equation}
\lambda_D = \left(\frac{k_BT_e\epsilon_0}{n_ee^2}\right)^{1/2} \approx 69\left(\frac{T_e}{n_e}\right)^{1/2} (\rm m).
\end{equation}
Here $k_B$ is Boltzmann's constant, $T_e$ and $n_e$ are the cloud temperature and density respectively, $\epsilon_0$ is the vacuum permittivity and $e$ is the unit electric charge. However, when filling the third stage, $\lambda_D \approx\ 1$ mm , i.e. the cloud is between the single particle and plasma regimes. When transferring into the HFT, the strength of the magnetic field, combined with the adiabatic transport of the positrons, will result in a radial compression of the cloud, such that $\lambda_D$ is now much smaller than typical cloud dimensions, and a non-neutral plasma is formed. The use of a cold trap ($\approx$ 10 K) cools the positrons down to 100 K or lower, thereby reinforcing plasma behaviour. For instance, charge screening alters the response to the external RW fields such that under the single particle conditions, the largest compression rate occurs at a drive frequency around that of the axial bounce \cite{Isaac11}, whilst for a plasma using the so-called strong drive regime, the density is proportional to the applied frequency \cite{PhysRevLett.94.035001,doi:10.1063/1.2179410}. 
 
The number of accumulated positrons, $N(t)$, at a loading time, $t$, can be described by the same simple exponential expression used for instruments loaded from radioactive source based beamlines as
\begin{equation} 
 N(t) = R\tau(1-e^{-t/\tau}),
 \label{eq:accucurve}
 \end{equation}
as long as the time between stacks is much smaller than the $\Pep$ lifetime $\tau$, with $R$ the trapping rate. Examples of such accumulation curves are given in section \ref{sec:First_and_Second_II}.

\subsection{Description of the equipment}
\label{BGT:Description}

\begin{figure*}[ht]
\centering
\includegraphics[width=1\textwidth]{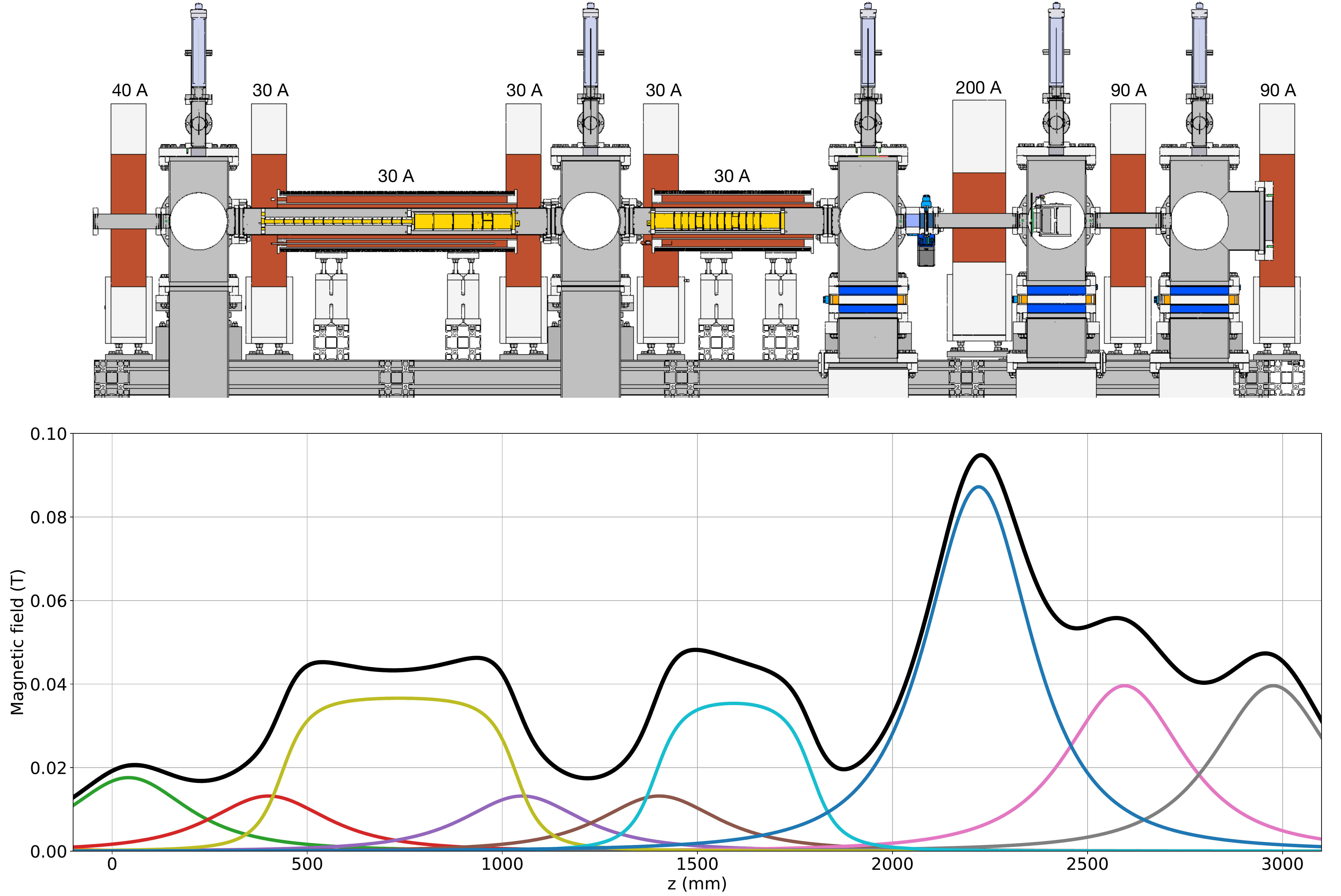}
\caption{Cross section of the three stage buffer gas trap (top) where the upstream side is on the left. There are two trapping areas, the first two stages (yellow, left) of the trap are inserted in a vacuum tube which is inside a solenoid between the first and second pumping stages, and similarly the third stage (yellow, right) is positioned between the second and third pumping stations. There are seven coils providing the magnetic field between the trap areas to guide the positrons. The numbers above the coils and the solenoids are the currents used to power the magnet. 
The calculated magnetic field along the axis of the system is shown in the bottom figure. The black line is the total field while the other coloured lines are the fields originating from the individual coils and solenoids.}
\label{fig:BGT}
\end{figure*}

The GBAR BGT is based on a system developed at Swansea University \cite{RevSciInst.77.063302} and has been built and tested at CEA-Saclay (see thesis Leite \cite{Amelia_Thesis}). As can be seen in figure \ref{fig:BGT}, it is a three stage accumulator, with the first two stages located between the first and second pumping stations (the pumps are connected at the bottom, but are not shown) separated from the third stage, which is situated between the second and third stations. There is a pumping restriction in the second pumping station. The first stage (see also inset in figure \ref{fig:TwoStagePot}) is comprised of 15 electrodes with an inner diameter of 16 mm and a length of 24 mm, directly connected to the second stage, which has 5 electrodes each with an inner diameter of 41 mm and a length of 49 mm. The fourth of these electrodes is cut into two axial pieces, with one separated azimuthally into 4 parts, for use as rotating wall electrodes. A 10 mm diameter, 200 mm long, aluminium pumping restriction has been placed in the second cross/pumping station. The third stage has 14 electrodes, also with an inner diameter of 41 mm. The length of each of these electrodes is chosen to be 17.4 mm, which is about 0.9 times the inner radius, thereby facilitating the formation of harmonic wells in almost any position in this region using a sequence of five electrodes \cite{NMnote}. The two outside electrodes on each side of the third stage electrode stack are 50 mm long.

The first two pumping crosses (see figure \ref{fig:BGT}) are each evacuated using an Oerlikon MAG W 600 turbomolecular pump, backed together by an Edwards XDS10 scroll pump. The third and fourth crosses are each serviced by SHI Cryogenics Marathon CP-8 cryopumps and the fifth cross by a SHI Cryogenic APD-8 cryopump. The vacuum tubes located between the third and fourth and the fourth and fifth crosses have an inner diameter of 20 mm and act as pumping restrictions. The base pressure in all five crosses is in the low $10^{-10}$ mbar range, measured using Pfeiffer cold cathode gauges. The nitrogen trapping gas is inserted in the middle of the first stage as illustrated schematically in figure \ref{fig:TwoStagePot}, whilst the carbon dioxide cooling gas is directly inserted into the second cross. The influx of the gases is regulated by HORIBA STEC digital Mass Flow Controllers SEC-Z502MG, calibrated for the gas used. The vacuum system is controlled using a National Instruments NI-cRIO-9066 (CompactRIO) system which is internally divided into realtime and field programmable gate array  parts. The system uses a number of digital input and output modules, together with serial interface modules, and is programmed using LabView. 

The BGT electrodes are immersed in magnetic fields produced by two home-built solenoids which provide a field of about 40 mT. Seven coils are installed to guide the positrons when moving between the traps. The currents to power the coils and solenoids originate from a number of Delta Elektronika 1500 W and 3000 W DC power supplies. The various contributions of the coils to the overall axial magnetic field are shown in figure \ref{fig:BGT}.

The experiment itself is controlled using a National Instruments PXIe-8135 computer/chassis. The potentials are initially produced by three 8-channel analog output PXI 6733 modules, with 740 kHz sampling rate and a range of -10 to 10 V, which subsequently feed amplifiers built by Aled Isaac from Swansea University, giving a range of $\pm 140$ V for each electrode. A bespoke control program steers the voltages, and uses a NI PXIe-7820R FPGA module to trigger switches, oscilloscopes (such as the 4-channel 2.5 GS/s PXIe 5160 and the 2 MHz analog input on the PXIe-6366 module) and more. Movable multi-channel plates (MCP) with phosphor screens are mounted in the first and fourth crosses and a metal plate to measure charge is mounted in the third cross. The pictures produced on the phosphor screen can be observed using cameras outside the vacuum chamber.

The HFT \cite{OSHIMA2003178,Grandemange_Thesis} is situated inside a 5 T superconducting magnet manufactured by Toshiba. Soft iron mounted close to the magnet vessel helps to provide a more homogeneous field inside the magnet with a  uniformity of the field  better than $10^{-3}$ inside a volume with radius of 2 mm and a length of 500 mm. The electrode stack consists of 27 electrodes with an inner diameter of 38 mm. The central 21 electrodes have a length of 22 mm and are enclosed by 160 mm long electrodes, while two 270 mm long outer electrodes on each side act as end caps. The total length of the assembly is 1.88 m (including insulating spacers).

The electrodes are powered in a similar way to the BGT, however with amplifiers that can deliver voltages between $\pm 4000$ V; such high voltages are needed to trap around $10^{10}$ positrons. The maximum absolute difference between consecutive electrodes is  $2000$ V. Two of the inner electrodes are divided azimuthally into four parts so they can be used to produce a rotating wall electric field. The superconducting magnet is cooled using a Sumitomo RDK-408D2 cold head, while a second cold head of the same type is connected to a copper tube in which the electrodes are installed, reaching a temperature of about 10 K. The vacuum system is pumped via an Agilent VacIon Plus 300. A moveable electron gun is installed on the upstream side of the magnet, while on the downstream side a moveable MCP can be used to diagnose (e.g., the size and density) the lepton clouds upon ejection from the trap.

\subsection{Effect of the electron beam on the positron accumulation and lifetime in the BGT}
\label{sec:First_and_Second}
\begin{figure*}[ht]
\centering
\includegraphics[width=0.9\textwidth]{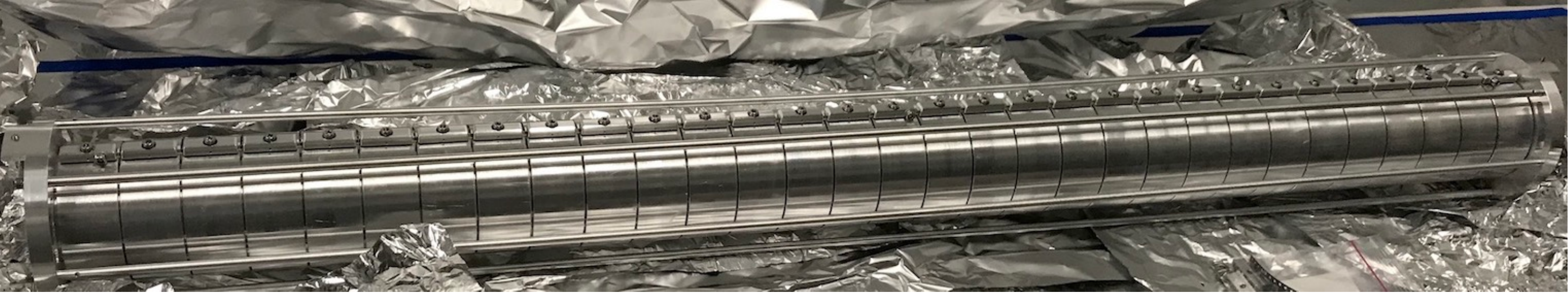}
\caption{The electron repeller consisting of 31 electrodes.}
\label{fig:repeller2}
\end{figure*}

The presence of electrons in the positron beam originating from the linac, as described in section \ref{sec:PosProd}, has a detrimental effect on the accumulation efficiency (see figure \ref{fig:Repeller_results}). The origin of this effect is not yet known, although one could speculate that their charge disrupts the positron capture. Initially, a tungsten mesh able to be biased up to 5000 V was inserted to reduce the electron flux. Unfortunately, the high fields around the grid wires had a marked effect on the adiabatic behaviour of positrons traversing the mesh, increasing the parallel energy width of the positron beam and so reducing the trapping efficiency. Also, the grid has a transparency of 90\%, commensurately reducing the number of positrons through annihilation. Therefore, a so-called electron repeller was developed, and inserted just before the entrance of the BGT. It consists of 31 electrodes, each with an inner diameter of 61 mm and a length of 28.4 mm, as shown in figure \ref{fig:repeller2}. The electrodes are connected to one another via 4.7 M$\Omega$ resistors. The central electrode is connected to a $-5$ kV power supply, whilst the two outer electrodes are grounded: 
\begin{figure}[ht]
\centering
\includegraphics[width=0.7\textwidth]{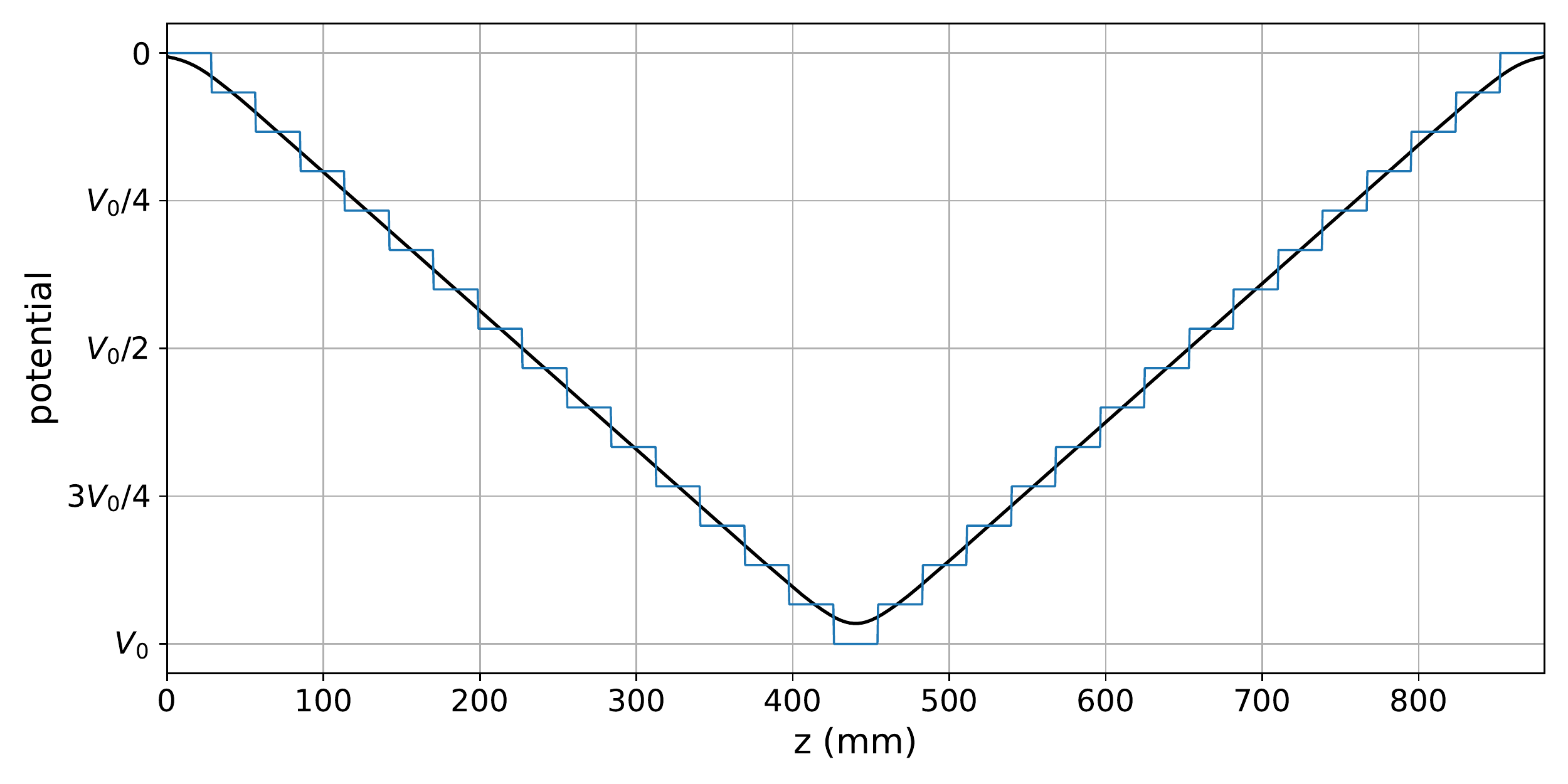}
\caption{Calculated electrical potential profile inside the repeller, at the electrodes radius (blue) and on axis (black). $V_0$ is the potential on the middle electrode.}
\label{fig:Repeller_Voltages}
\end{figure}
the resulting potential inside the repeller is shown in figure \ref{fig:Repeller_Voltages}. The  axial potential change is very smooth so that the positrons, with initial energy of about 50 eV, keep the same parallel energy width across the repeller region. 

In figure \ref{fig:Repeller_results} the number of electrons and the parallel energy width of the positron beam together with the trapping rate are plotted as a function of the applied repeller voltage. It is clear that the positrons move adiabatically through the repeller region, while the trapping rate increases with increasing repeller voltage, stabilising at around 1000 V. The electrons do not seem to have an effect on the lifetime of the positrons, however in the second stage of the BGT the latter is mainly limited by the number of cooling and trapping gas atoms. It is not yet clear if electrons with energies $> 5$ keV coming from the linac will have an effect on the properties of a positron plasma in the HFT.
\begin{figure}[htbp] 
   \centering
   \includegraphics[width=\linewidth]{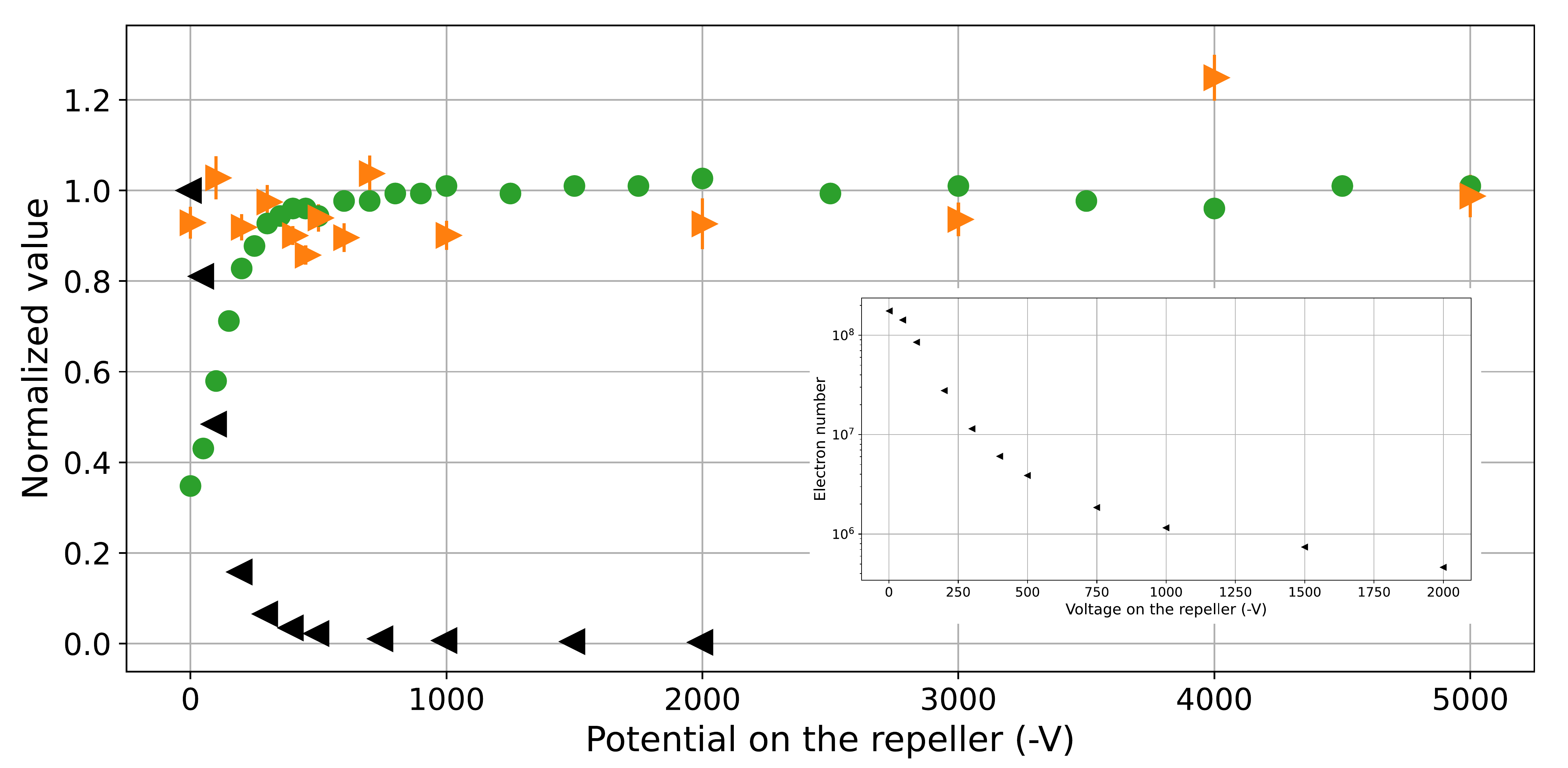} 
   \caption{Normalised values for the number of electrons ($\blacktriangleleft$), positron trapping rate ({\color{limegreen}$\bullet$}) in the second stage of the BGT and the parallel energy width of the positrons ({\color{saffron}$\blacktriangleright$}). The RW was on during the accumulation using a frequency of \SI{2.4}{\mega\hertz} and a voltage on the electrodes of \SI{1}{\volt}. Inset: number of electrons on a log-lin scale. The linac frequency was \SI{200}{\hertz}.}
   \label{fig:Repeller_results}
\end{figure}

\subsection{Energy width of the positrons}
\label{subseq:EnerWidthPos}
The parallel energy width of the positron beam was measured just after the moderator, where the magnetic field is 9.7 mT, using potentials on a grid. The number of transmitted positrons was determined using a CsI detector. The resulting data have been fitted with a  complementary error function (erfc), where a standard deviation, $\sigma$, of about 2.5 eV gave the best fit results. Subsequently the positrons move adiabatically into the BGT, where the magnetic field is about 45 mT. Here the measurement of the parallel energy width was performed using a number of electrodes to provide a flat potential in their centre, and a $\sigma$ of 4.5 eV was found. At low parallel energies the positrons may not be able to enter the trap. Increasing the moderator magnetic field would give a better trapping rate of the incoming positrons, however, the inner electrode radius is presently too small and  positrons are lost because the radial extent of the beam increases beyond the radius of the trap when the moderator magnetic field becomes higher than 9.7 mT. The implication is that only $2.6\times 10^7$ positrons per second enter the BGT, about 65\% of the $4\times10^7$ e$^+$s$^{-1}$ emanating from the linac when operating at 200 Hz.

\subsection{Initial trapping in the first and second BGT stages}
\label{sec:First_and_Second_II}

\begin{figure*}[ht]
\centering
\includegraphics[width=0.8\textwidth]{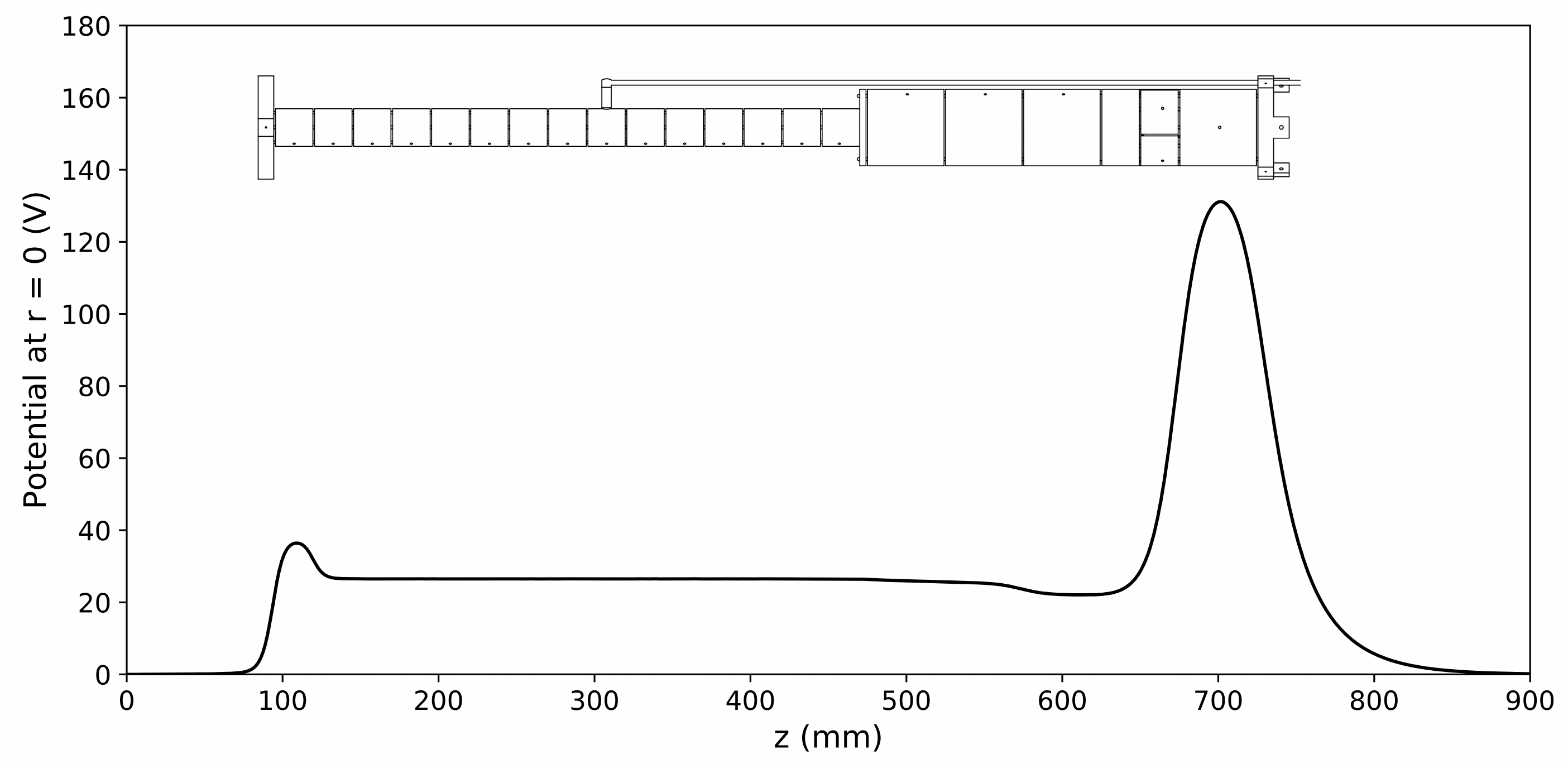}
\caption{Schematic illustration of the first two stages of the BGT, together with the calculated axial trapping potential on the z-axis, in the centre of the electrodes. The last but one electrode, going from left to right, is segmented in four parts and this is where the four RW voltages are applied. Nitrogen gas is injected into the middle of the first stage using the delivery pipe shown, whilst carbon dioxide is inserted into the second vacuum cross (see figure \ref{fig:BGT}) and subsequently diffuses into the second and third stages.}
\label{fig:TwoStagePot}
\end{figure*}

A schematic diagram of the electrodes used in the first two stages of the BGT is shown in figure \ref{fig:TwoStagePot}, together with the trapping potential. Nitrogen gas is inserted into the central electrode of the first stage while in between the second and third stages CO$_2$ molecules are added (when required). Incoming positrons that excite the nitrogen molecules as described by equation (\ref{eq:ElecExcite}) will be trapped between the first and last electrodes. A subsequent N$_2$ excitation confines the positrons in the second stage, where the pressure is about ten times lower than in the first stage. As described further in section \ref{sec:RWoptimisation}, a 2 MHz rotating wall electric field with an amplitude of 1 V applied using the segmented electrode shown in figure \ref{fig:TwoStagePot} is present during the capture and hold, and as such compresses the accumulated positron cloud.  The heating of the positrons due to the RW is counteracted by cooling collisions on the carbon dioxide. 
\begin{figure}[htbp] 
   \centering
   \includegraphics[width=.5\linewidth]{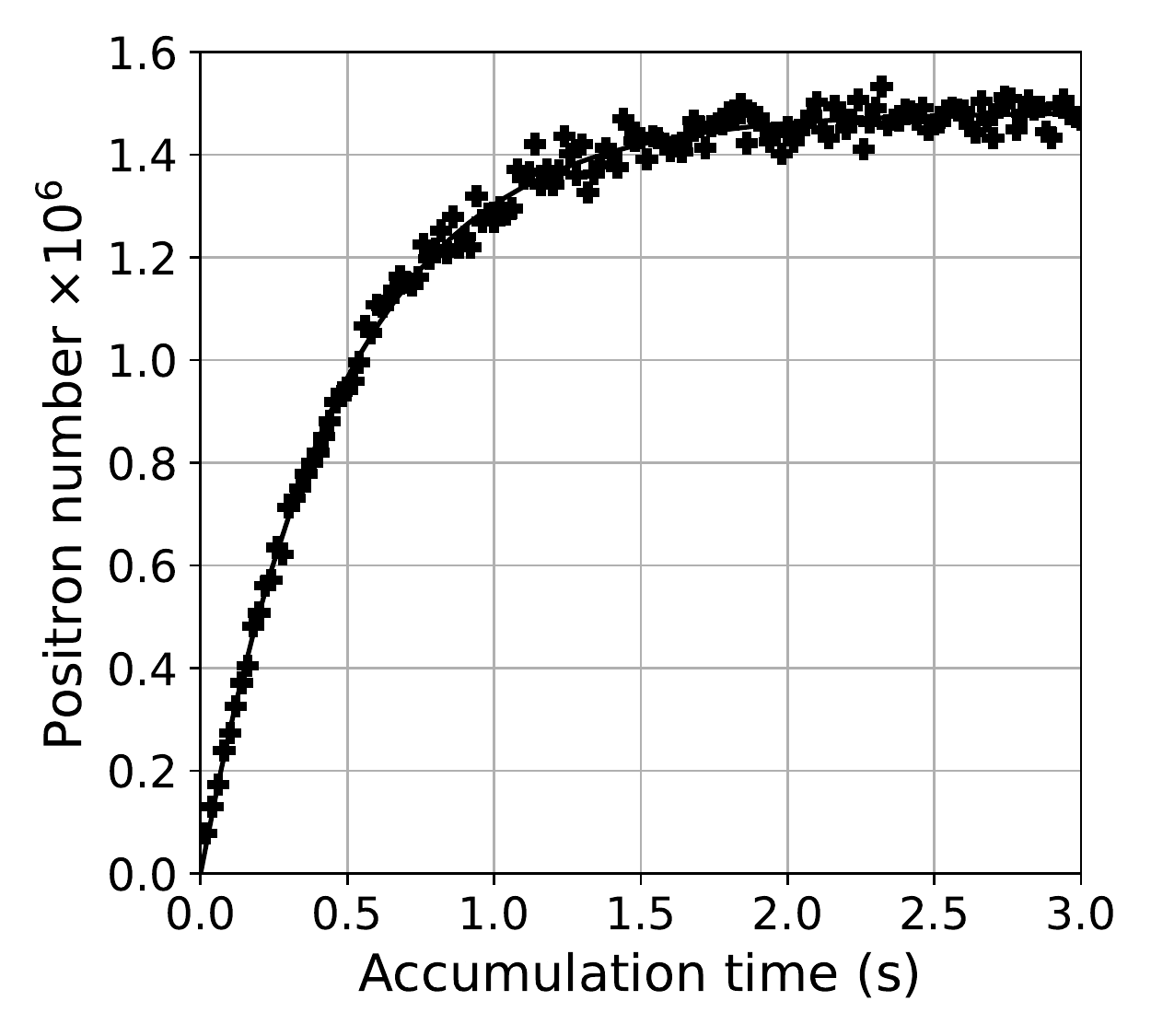} 
   \caption{Positron accumulation in the second stage. RW parameters: \SI{2}{\mega\hertz}, \SI{1}{\volt}. Fitting the measured values to equation (\ref{eq:accucurve}) yielded $\tau = \SI{0.47 +- 0.001}{\second}$ and $R = \SI{3.1 +- 0.06e6}{\per\second}$.}
   \label{fig:Accu1}
\end{figure}

Figure \ref{fig:Accu1} presents a so-called accumulation curve for the first two stages, using a second stage nitrogen pressure of $\sim 10^{-4}$ mbar and a carbon dioxide  pressure of $\sim10^{-5}$ mbar. We used an accumulation time of 100 ms, where the accumulation still follows a linear behaviour with time, for stacking into the third stage (see section \ref{subsec:Third_BGT}).

\begin{figure}[ht]
\centering
\includegraphics[width=0.9\textwidth]{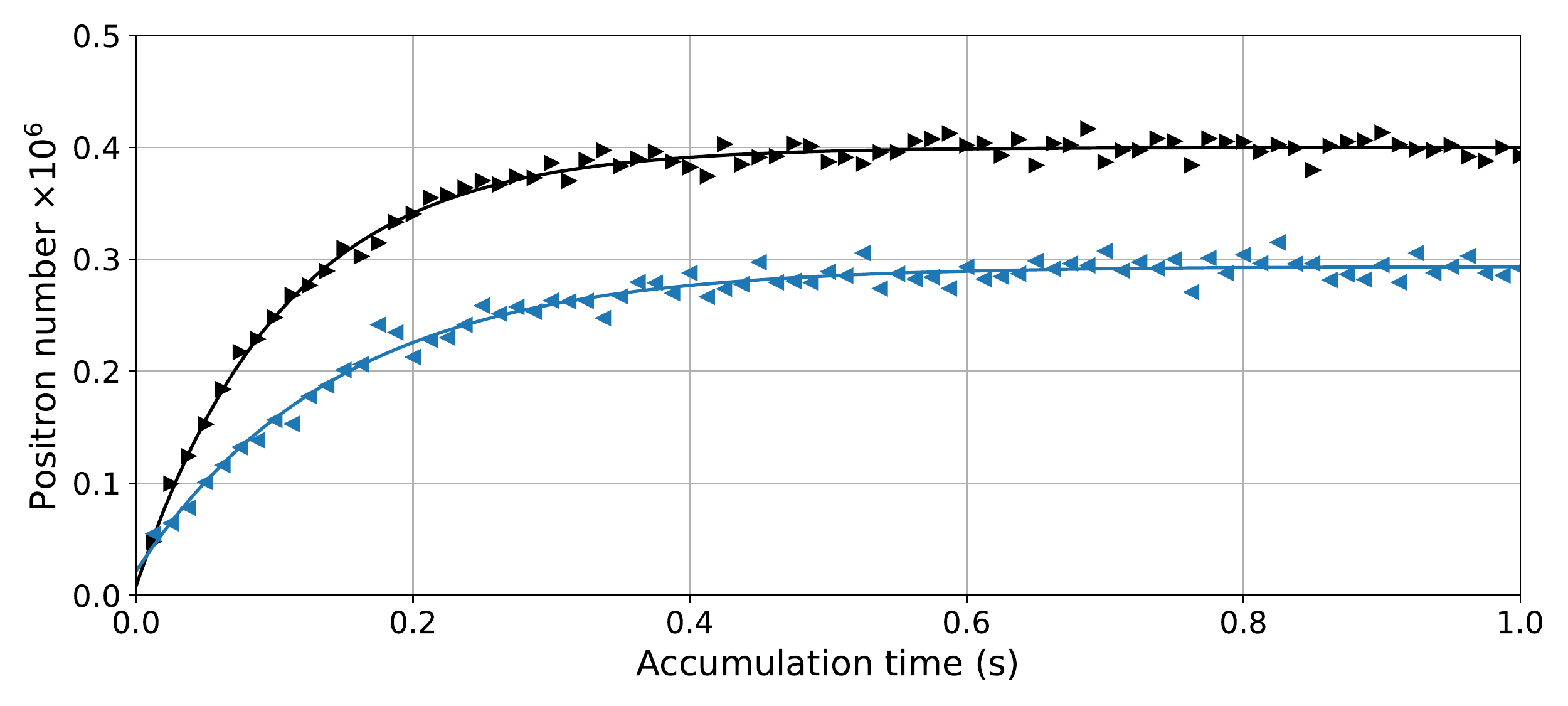}
\caption{Accumulation in the second stage without using the RW. The blue points ({\color{Cerulean}$\blacktriangleleft$}) are measurements taken while there was only nitrogen gas inserted into the system. The  fitted line gives $\tau = \SI{0.144+-0.006}{\second}$ and $R = \SI{1.9+-0.1e6}{\pos\per\second}$. The black points ($\blacktriangleright$) were taken with both nitrogen and carbon dioxide  present and with fitted values of $\tau = \SI{0.105+-0.003}{\second}$ and $R = \SI{3.7+-0.2e6}{\pos\per\second}$.}
\label{fig:Compare_wwoRW_wwwCO2}
\end{figure}

The positron trapping rate (not using the RW), when adding CO$_2$ gas to the nitrogen, shows a marked increase (see figure \ref{fig:Compare_wwoRW_wwwCO2}). The positron trapping probability of CO$_2$ is 16\% of that for nitrogen at the same pressure \cite{GREAVES200290}, the pressure ratio of CO$_2$/N$_2$ in the first stage where the majority of the trapping occurs at
 about 0.01, so the observed effect was not expected. However, the figure shows an increase of about 40\% when using both gases compared with the results of nitrogen only.  It seems that we cannot explain the increased positron trapping rate by direct excitation, nor through the vibrational energy loss channels \cite{10.1088/1361-6455/aba10c}. A possible explanation is that the CO$_2$ molecules improve the transfer efficiency between the first and the second stage, but further measurements are needed to investigate this effect.

\subsection{Rotating wall optimisation}
\label{sec:RWoptimisation}
\begin{figure*}[htbp]
  		\centering
  		\includegraphics[width=\linewidth]{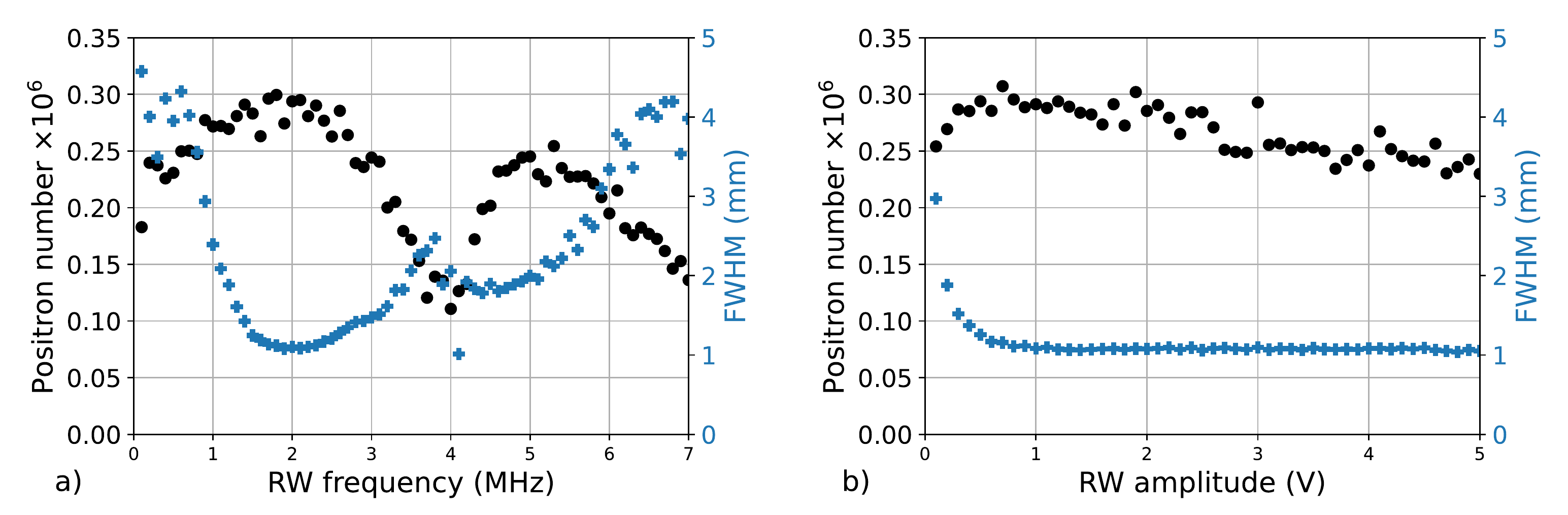}
		\label{fig:RW_amp_2nd}
	\caption{Positron number ejected from the BGT's second stage ({\fontsize{10pt}{\baselineskip}$\bullet$}) and FWHM of the positron cloud imaged on the MCP ({\color{Cerulean}\fontsize{7pt}{\baselineskip}\ding{58}}) after \SI{100}{\milli\second} accumulation. a) as a function of RW frequency, RW amplitude: \SI{1}{\volt} b) as a function of the RW amplitude, RW frequency: \SI{2}{\mega\hertz}.}
\label{fig:RW_2}
\end{figure*}

As can been seen in figure \ref{fig:TwoStagePot}, the potential well is not harmonic. However, the positron cloud after 100 ms accumulation is still not a plasma, so the bounce frequency of the particles in the trap is, based on a simple calculation, about one MHz wide. This can be observed by the broad frequency range within which one is able to radially compress the positrons, as can be seen in figure \ref{fig:RW_2}a. The optimum RW frequency is about 2 MHz, with an amplitude of 1 V, found by varying the amplitude (figure \ref{fig:RW_2}b). The losses at higher frequencies and larger potentials are attributed to particle heating. Note that these values are chosen from a flat area in the graphs. Another method to decrease the cloud diameter is by applying a frequency chirp, i.e. changing the frequency from a higher to a lower value \cite{Deller_2014}, but this has not as yet been investigated.

\subsection{Transfer into the third BGT stage}
\label{subsec:Third_BGT}
\begin{figure}[htbp] 
   \centering
   \includegraphics[width=.75\linewidth]{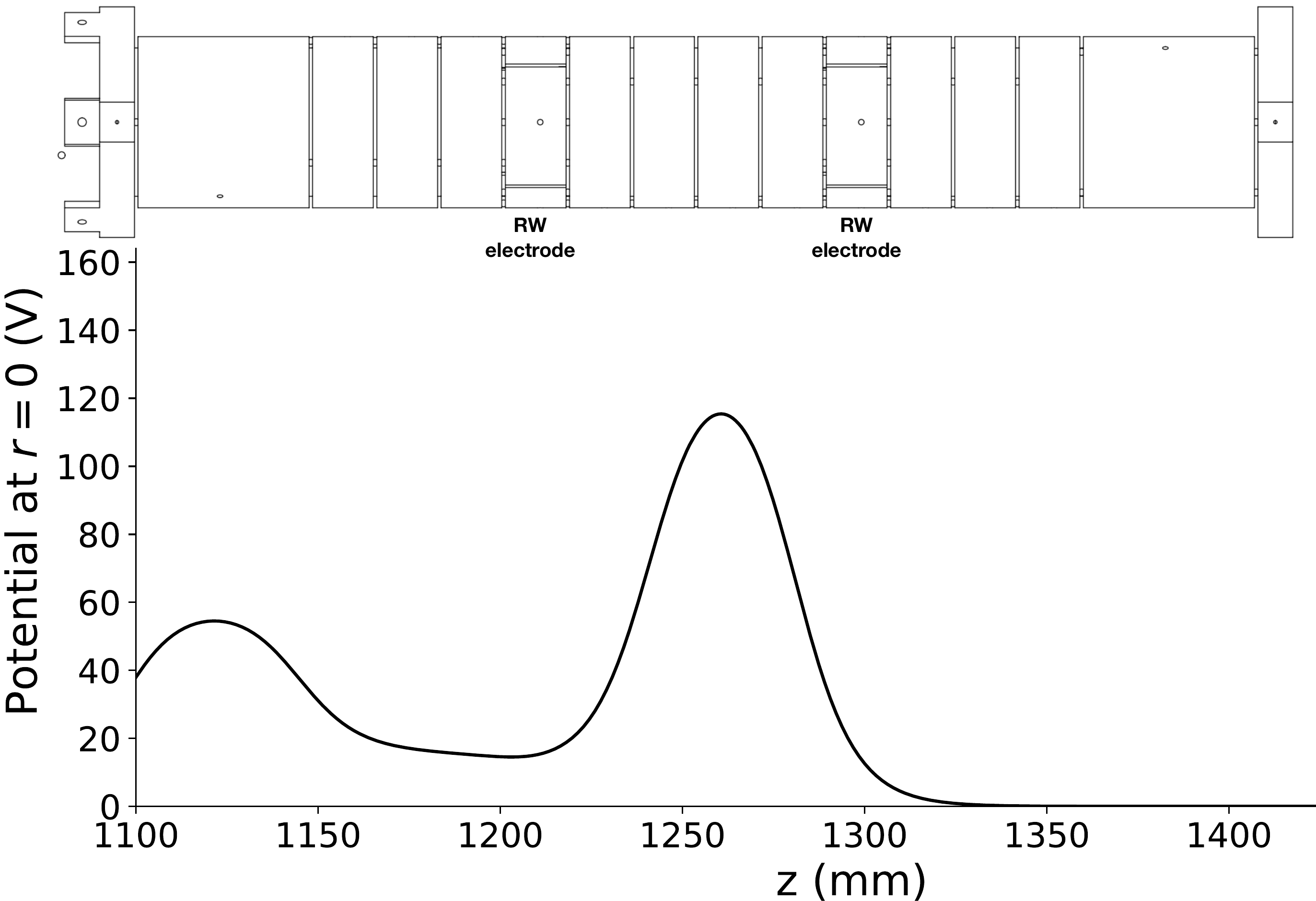} 
   \caption{Schematic illustration of the third BGT stage together with the axial (along $z$, with $r=0$) trapping potential used.}
   \label{fig:potential_3rd}
\end{figure}
While $\Pep$ accumulation into the first two BGT stages is still roughly proportional to the accumulation time (for instance 300 ms for the accumulation curve in figure \ref{fig:Accu1}), the RW is switched off at 100 ms accumulation, whereafter the positron cloud is axially compressed and then ejected from the second stage by quickly ($\sim$ 30 ns) lowering the potential applied to the last electrode of that stage, and then re-trapped in the third stage (see figure \ref{fig:potential_3rd}). We refer to this as a stack. 
\begin{figure*}[htbp]
  		\centering
		\includegraphics[width=\linewidth]{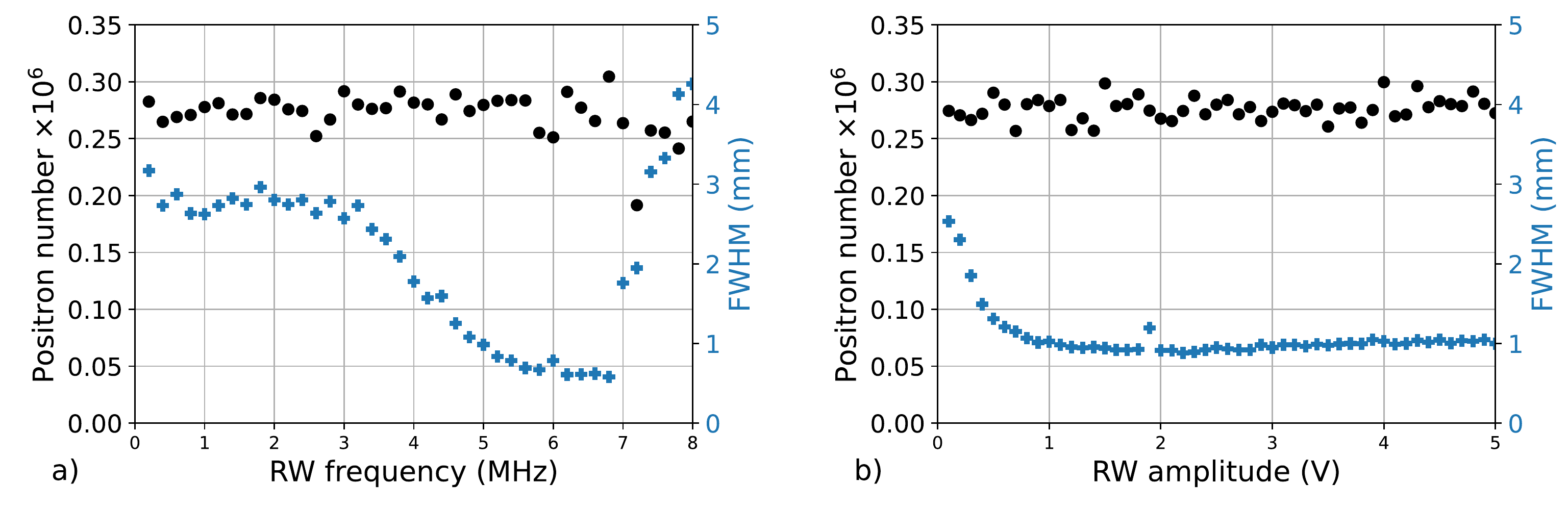}
	\caption{(a) Positron number ejected from the BGT's third stage ($\bullet$) and full width at half maximum of the positron cloud ({\color{Cerulean}\fontsize{7pt}{\baselineskip}\ding{58}}) a) as a function of the RW frequency, with RW amplitude: \SI{1}{\volt} and b) as function of the RW amplitude with RW frequency: \SI{5}{\mega\hertz}.}
	\label{fig:RW_3}
\end{figure*}
A RW applied on the first segmented electrode in the third stage subsequently centres and compresses the particles using a drive frequency of 5 MHz (see figure \ref{fig:RW_3}a) and an amplitude of 5 V (see figure \ref{fig:RW_3}b). The measured particle density is not dependent on the number of stacks because the radius of the $z$-integrated positron cloud, as measured on the MCP, is proportional to the square root of the number of stacks. The temperature of the positrons is only dependent on the amount of gas in the system, which leads us to conclude that the Debye length does not change.
\begin{figure}[htbp] 
   \centering
   \includegraphics[width=.5\linewidth]{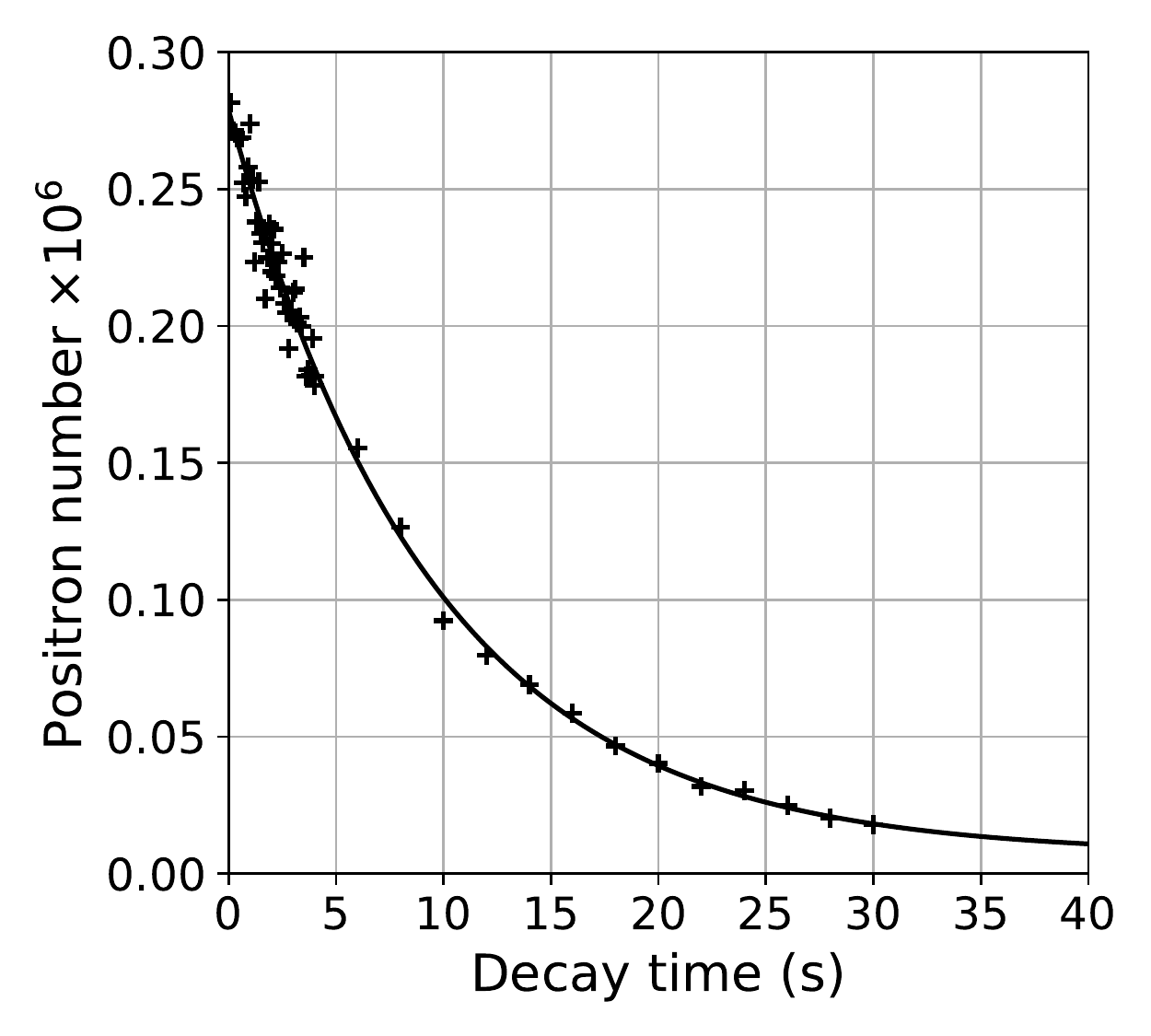} 
   \caption{Decay curve in the third stage. The bunch of positrons corresponds to \SI{100}{\milli\second} accumulation in the second stage. The measured points are fitted to $N(t) = N_0\exp^{-t/\tau}$ resulting in  $N_0 = \num{2.71+-0.05d5}, \tau = \SI{9.41 +- 0.57}{\second}$.}
   \label{fig:fig4-decay}
\end{figure}

The positron lifetime curve of one stack is plotted in figure \ref{fig:fig4-decay}. The fitted lifetime is about 10 s, long enough for stacking positrons for at least up to 1 second (10 stacks) without noticeable loss due to annihilation. Due to the pumping restriction between the second cross and the third stage, both the N$_2$ and CO$_2$ pressures are reduced, resulting in a factor of about 20 increase in the  lifetime of the positrons. 
\begin{figure*}[htbp]
  		\includegraphics[width=\linewidth]{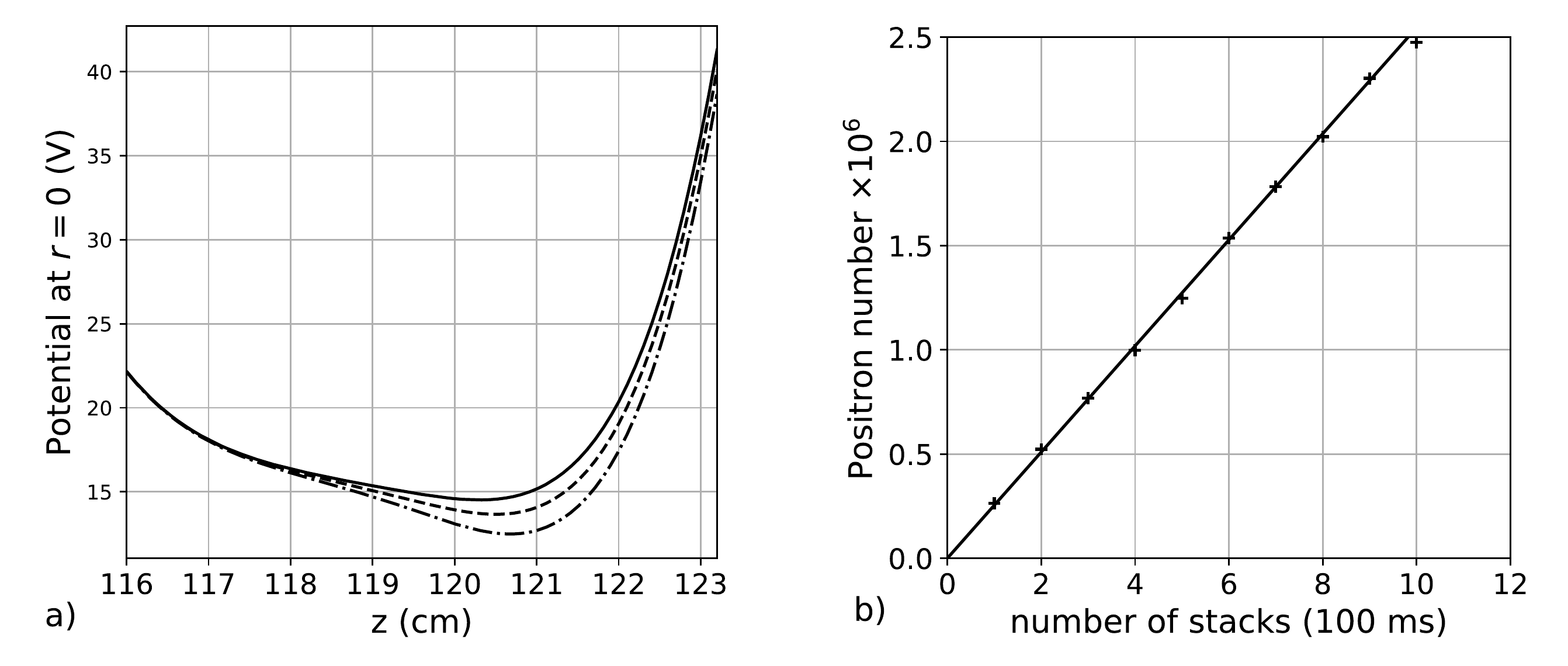}
	\caption{a) Potential profiles in the third stage used for the stacking procedure. For each stack, the bottom of the well is decreased by \SI{0.4}{\volt} a) Potentials for the first stack (solid line), the fifth stack (broken line) and the tenth stack (dash-dot line). b) Positron number as a function of the number of stacks.}
    \label{fig:fig5-third}
\end{figure*}

The efficiency of re-trapping the positrons depends on the value of the well depth, so after each stack the potential minimum is lowered by 0.4 volt (see figure \ref{fig:fig5-third}a). In figure \ref{fig:fig5-third}b the filling of the well is shown, from which we conclude that up to 10 stacks there is a linear behaviour with a rate of about $\SI{2.55+-0.01e6}{\pos\per\second}$, indicating a transfer efficiency from the second to third stage of 80\%, and a total efficiency compared with the incoming beam strength of $2.6\times 10^7$ s$^{-1}$ of about 10\%. A recent publication from an experiment based in Hiroshima showed that when using a linac combined with a three stage accumulator, an efficiency of 4\% has recently been reached \cite{Higaki_2020} and a trapping rate of $1.7\times 10^5$ \Pep s$^{-1}$ has been obtained.

\section{High Field Trap}
\label{sec:HFT}
Once the third stage has accumulated 10 stacks, the positrons are transferred into the HFT using a similar method as applied when moving the positrons between the second and third stages. The particles in the BGT are axially compressed before the last electrode is lowered to zero volt in about 30 ns. Before that, the entrance to the HFT is lowered, however there is still a barrier so that the positrons which are already inside and have been cooled by synchrotron radiation cannot escape (at 5 T the cooling time is about 0.16 s). When the whole bunch is inside the trap the entrance electrode potential is quickly raised so the positrons cannot escape.  Thereafter the RW is switched on to compress the new stack and merge it with the already trapped particles.
\begin{figure}[ht]
\centering
\includegraphics[width=0.5\textwidth]{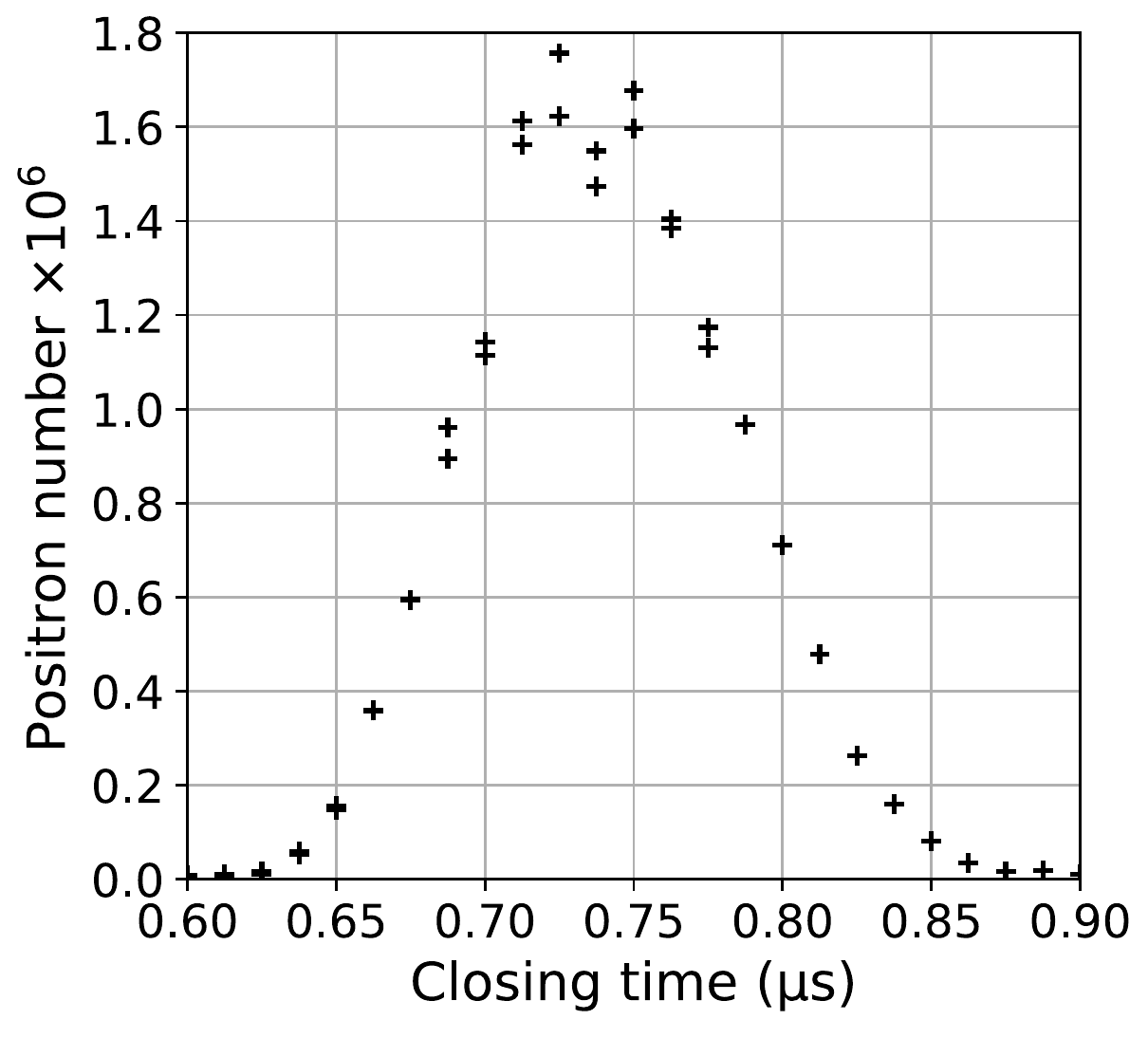}
\caption{Number of positrons trapped as function of closing time of the entrance of the HFT.}
\label{fig:Capture_in_HFT}
\end{figure}

Figure \ref{fig:Capture_in_HFT}  shows the number of trapped positrons as a function of the closing time of the trap, following the opening of the last electrode of the third stage. There is a clear maximum between 700 and 750 ns. The resulting positron cloud for one stack showed two different lifetimes; a short component with a lifetime of less than a second, and one with a lifetime between 100 and 1000 seconds. The relative fraction and lifetimes of the two populations were shown to change with the well depth, but this is not yet well understood \cite{Samuel_Thesis}.

\begin{figure}[ht]
\centering
\includegraphics[width=0.5\textwidth]{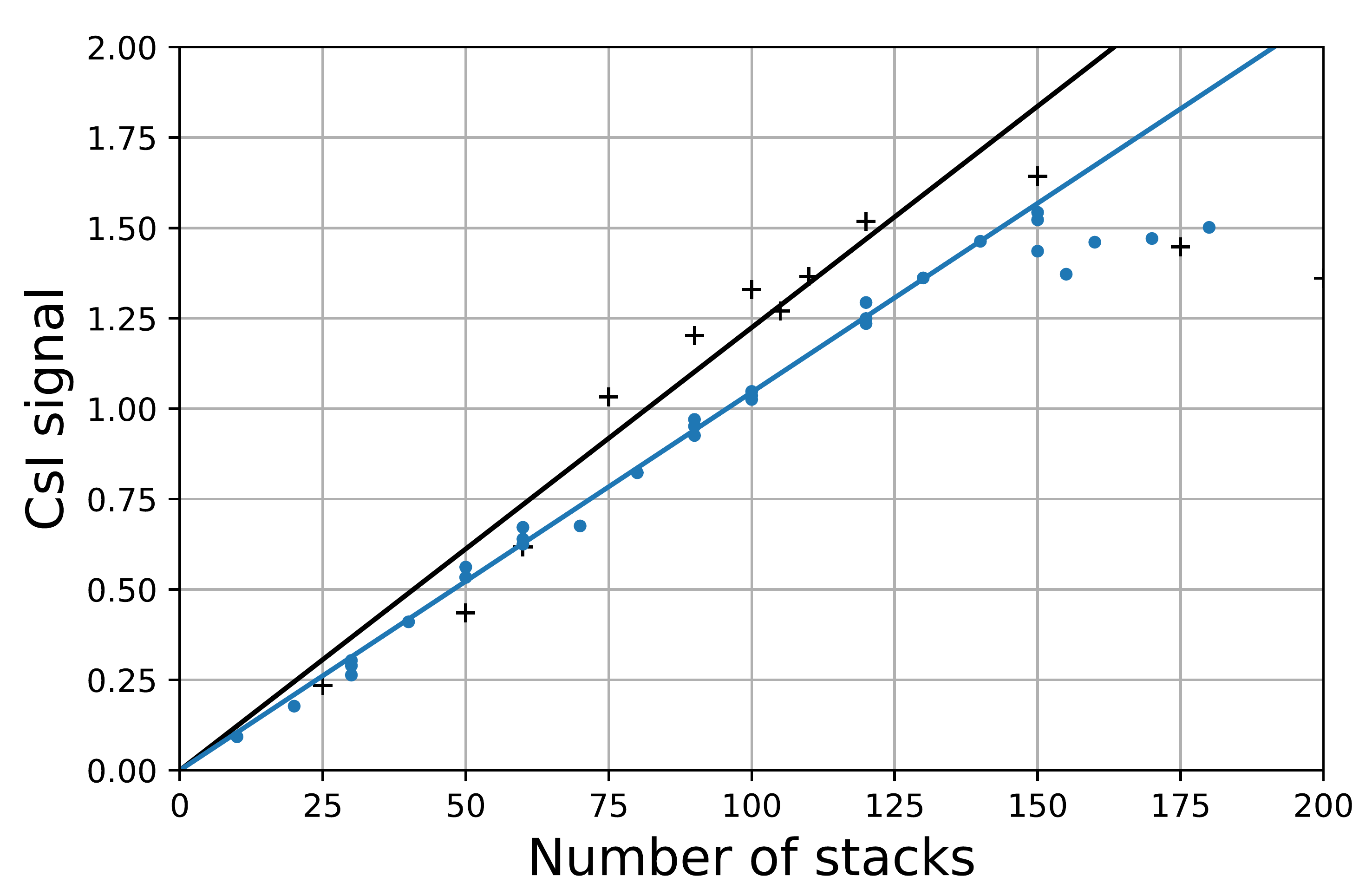}
\caption{Measurement of the effect of the rotating wall on the trapping efficiency in the HFT. a) Black crosses (RW on) and blue dots (RW off). The lines are a linear fit to the points up to the CsI detector value (proportional with the number of positrons) of 1.5.}
\label{fig:HFT_RW_or_not_RW}
\end{figure}
Just as in the third stage of the BGT, we also stack in the HFT, but in this case every 1.1 seconds. The resulting accumulation curve for the first 175 stacks is shown in figure \ref{fig:HFT_RW_or_not_RW}, where we clearly can see that the RW has a positive influence on the trapping rate in the HFT  after 70 stacks.
\begin{figure*}[htbp]
  		\centering
  		\includegraphics[width=\linewidth]{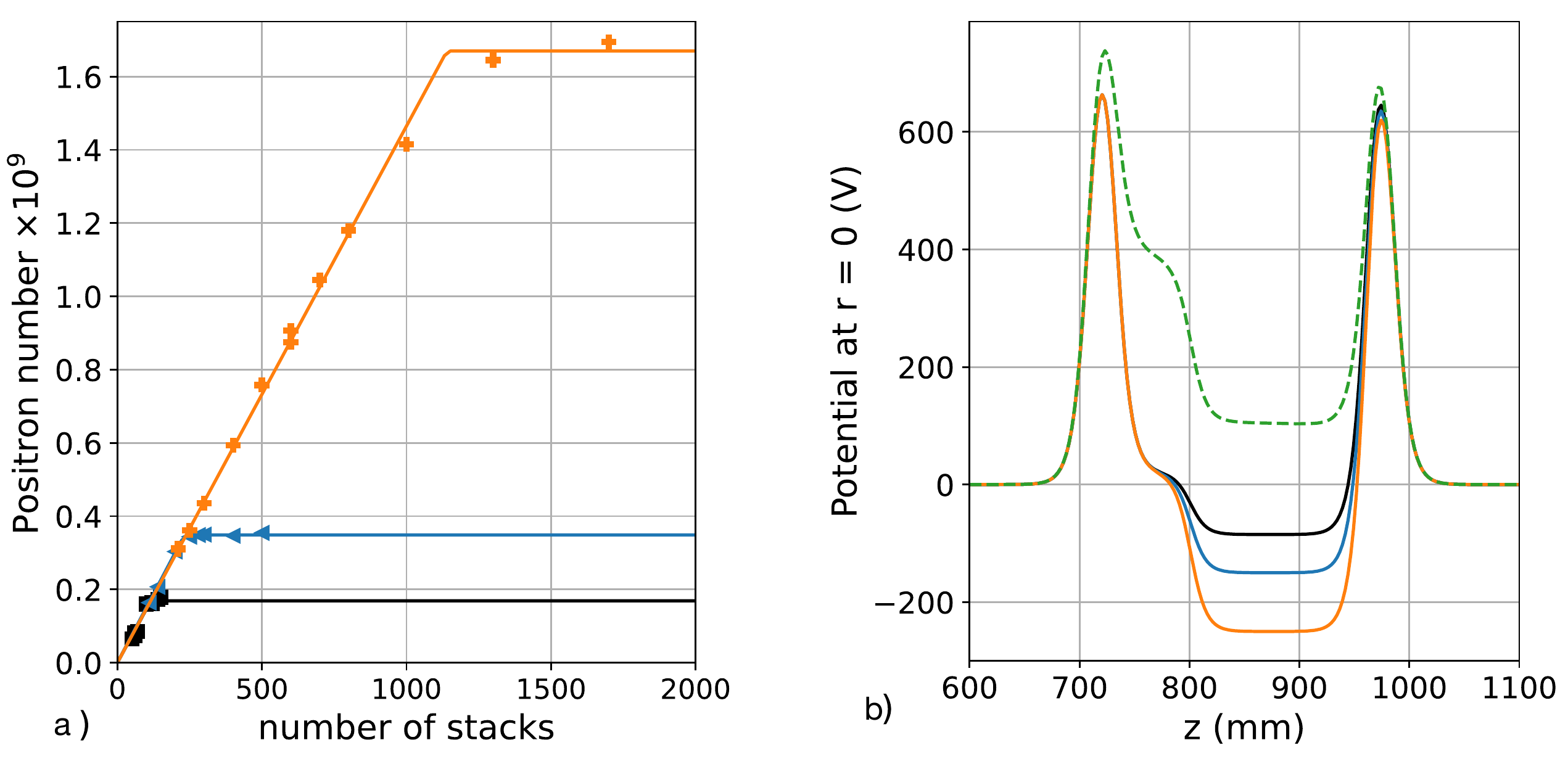}
  	\caption{a) Positron number as a function of the number of stacks into the HFT for the successive potential wells. The colours of the lines correspond to the colours plotted for the different wells shown in b). b) Potential profiles used for the stacking (solid lines) and the ejection (broken line) procedures. }
		\label{fig:stacking_HFT}
\end{figure*}
As in trapping in the third BGT stage, after a certain number of stacks the well is full and the next stack will no longer be trapped, as shown in figure \ref{fig:stacking_HFT}a. Thus, the bottom potential is lowered by a certain amount, dependent on how many stacks are trapped (see figure \ref{fig:stacking_HFT}b). For 1000 stacks we were able to trap  \num{1.4+-0.2e9} positrons in 1100 seconds. Compared to the positron beam entering into the BGT, we have an overall trapping efficiency of about 5\%.  

\section{Conclusions and outlook}
We have built and commissioned a Buffer Gas Trap followed by a High Field Trap able to accumulate \num{1.4+-0.2e9} positrons in 1100 seconds, a record trapping rate. The ultimate goal is to obtain $3\times 10^{10}$ positrons in about 100 seconds, instead of the $1.36\times 10^8$ presently available (5\% of the number of incoming positrons in 100 s). There are a number of changes/optimisations which could improve the number of positrons trapped in the HFT. These range from improving the flux of positrons emanating from the linac, improving trapping rates in the BGT and enhancing the lifetime of the trapped plasma in the HFT, as listed below:
\begin{itemize}

\item  Most measurements shown in this article are taken at a linac repetition rate of 200 Hz due to stability issues when operating the device at 300 Hz for long periods. This should be solved soon and will lead to a 25 \% increase in the positron flux. The number of positrons per pulse decreases with increasing linac frequency, this is likely due to the temperature increase of the moderator. Possible solutions have been proposed, such as moving the electron target beam slightly every pulse or have a rotating moderator. If this effect could be negated, a gain of 40\% could be expected. Furthermore, we only used 10 stacks in 1.1 seconds (every stack was accumulated in 100 ms) so there could be a gain of 10\%. Combining these factors should lead to a factor of 1.9 increase in the positron flux coming from the moderator.

\item The 6 \% trapping efficiency in the first stages of the BGT is low compared to similar BGTs using a $^{22}$Na source, together with a neon moderator. Although the parallel energy distribution directly after the moderator is similar in both cases, the size of the beam spot when using the linac is larger. To fit the whole beam inside the first stage (16 mm diameter), a higher magnetic field inside the first stage is needed with the consequent increase in parallel energy distribution leading to a reduction in the trapping efficiency as explained in section \ref{subseq:EnerWidthPos}. A solution would be to increase the size of the electrodes in the first stage. Reducing the target spot size would probably lead to too much heating, which may damage the tungsten target. Another possibility could be using a reflection re-moderator made from silicon carbide \cite{St_rmer_1996,doi:10.1063/1.1435838} in combination with a cooling gas. Preliminary experiments using this material are encouraging, and would give a higher trapping rate than is possible with nitrogen gas. \\

\item When \Pep ~are transferred into the HFT, only 75\% of them are re-trapped. It is not yet clear if the positrons hit the first electrode in the HFT,  if the loss is due to magnetic mirroring or the if 25\% of the positrons are lost immediately after entering the HFT. Moreover, after transfer of the first stack, two populations with quite different lifetimes are re-trapped. The short lifetime is around 0.3 s, so no \Pep~ from this population (around 25\% \cite{Samuel_Thesis}) remain after a second or so. More research is needed to get a better understanding of the transfer and trapping, and to find a method to produce only a long lifetime population. The lifetime of up to 40 minutes of the remaining positrons is still short, and hopefully will be improved by lowering the electrode temperatures, better pumping, and stopping the BGT-gas from entering the HFT.\\

\end{itemize}

If we are able to make the improvements above, $10^{10}$ \Pep could be accumulated in about 750 seconds. Although not yet the goal stated in the introduction, it should be sufficient to allow observations of the formation of \Hbarp via the reaction shown in equation (\ref{eq:Hbarp}).
  
\label{sec:conclusions}
\section*{\bf Declaration of competing interest}
The authors declare that they have no known competing financial interests or personal relationships that could have appeared to influence the work reported in this paper.

\section*{\bf Acknowledgments}
We acknowledge support of the Agence Nationale de la Recherche, France (project ANTION ANR-14-CE33-0008), CNES, France (convention number 5100017115), the Swiss National Science Foundation, Switzerland (Grant number 173597), ETH Zurich, Switzerland (Grant number ETH-46 17-1), the Swedish Research Council (VR), the Ministry of Education of the Republic of Korea and the National Research Foundation of Korea (NRF-2016R1A6 A3A11932936, NRF-2021R1A2C3010989). Laboratoire Kastler Brossel (LKB) is a Unit\'e Mixte de Recherche de Sorbonne Universit\'e, de ENS-PSL Research University, du Coll\`ege de France et du CNRS no 8552. We thank CERN for the support to construct the linac and its biological shield and for fellowships and Scientific Associateship provided to D. P. van der Werf and P. P\'erez. The support of the Enhanced Eurotalents Fellowship programme to D. P. van der Werf is acknowledged . We also thank CUP(IBS) for the fellowship of B. H. Kim (grant number IBS-R016-Y1). S. Niang acknowledges support from the Laboratoire dÕExcellence P2IO (ANR-10-LABX-0038) in the framework Investissements dÕAvenir (ANR-11-IDEX-0003-01). We gratefully acknowledge the help of Fran\c cois Butin and the CERN management and teams that were involved in this project, and are indebted to the technical support provided by Julian Kivell and Phil Hopkins from Swansea University and by Didier Pierrepont from CEA-Saclay.

\section*{\bf References}
\bibliographystyle{elsart-num}
\bibliography{GBAR_BGT}
\end{document}